\documentclass[aps,prl,twocolumn,superscriptaddress]{revtex4-2}

\usepackage[hidelinks]{hyperref}
\hypersetup{
	colorlinks,
	linkcolor={red},
	citecolor={blue},
	urlcolor={blue}
}
\usepackage{footnote}
\usepackage{physics}
\usepackage{balance}
\usepackage{amssymb}
\usepackage{suffix}
\usepackage{mathtools}
\usepackage[utf8]{inputenc}
\usepackage{booktabs}
\usepackage{cases}
\usepackage[multiple]{footmisc}
\usepackage{dcolumn}
\usepackage{color,soul}
\usepackage{rotating}
\usepackage{perpage}
\usepackage{siunitx}
\usepackage{xcolor}
\usepackage{soul}
\usepackage{amsmath}
\usepackage{tikz}
\usepackage[T1]{fontenc}
\usepackage{etoolbox}
\usepackage{graphics}
\usepackage{siunitx}
\usepackage{float}	
\usepackage{collref}
\usepackage{multirow}
\usepackage{mathtools}
\usepackage{bm}
\usepackage{url}

\usepackage{tikz}
\usepackage{tikz-3dplot}
\usepackage{accents}

\makeatletter
\newcommand{\doublewidetilde}[1]{{%
		\mathpalette\double@widetilde{#1}%
}}
\newcommand{\double@widetilde}[2]{%
	\sbox\z@{$\m@th#1\widetilde{#2}$}%
	\ht\z@=.9\ht\z@
	\widetilde{\box\z@}%
}
\makeatother

\usepackage{etoolbox,lipsum}

\newcommand{\blue}[1]{\textcolor{blue}{#1}}

\usepackage{braket}
\usepackage{physics}

\begin{document}
	\title{Cavity-induced coherent magnetization and polaritons in altermagnets}
	\author{Mohsen Yarmohammadi}
	\email{mohsen.yarmohammadi@georgetown.edu}
	\address{Department of Physics, Georgetown University, Washington DC 20057, USA}
	\author{Libor \v{S}mejkal}
	\email{lsmejkal@pks.mpg.de}
	\address{Max Planck Institute for the Physics of Complex Systems, N\"othnitzer Str. 38, 01187 Dresden, Germany}
	\address{Max Planck Institute for Chemical Physics of Solids, N\"othnitzer Str. 40, 01187 Dresden, Germany}
	\address{Institute of Physics, Czech Academy of Sciences, Cukrovarnick\'a 10, 162 00 Praha 6, Czech Republic}
	\author{James K. Freericks}
	\email{james.freericks@georgetown.edu}
	\address{Department of Physics, Georgetown University, Washington DC 20057, USA}
	
	\begin{abstract}
		Altermagnets feature antiparallel spin sublattices with $d$-, $g$-, or $i$-wave spin order, yielding nonrelativistic spin splitting without net magnetization. We show that embedding a two-dimensional $d$-wave altermagnet in a driven optical cavity induces a finite, tunable magnetization. Coherent photon driving couples selectively to electronic sublattices, and the resulting altermagnets’ symmetry-broken spin texture yields a pronounced steady-state spin imbalance---coherent magnetization--- absent in conventional antiferromagnets for the same lattice configuration. A mean-field Lindblad analysis reveals the dominance of quadratic over linear couplings. In the strong-coupling regime, distinct polariton signatures emerge in the steady state of induced magnetization. This work demonstrates cavity control of altermagnets for spintronic applications.
	\end{abstract}
	
	\maketitle
	{\allowdisplaybreaks	
		\blue{\textit{Introduction.}}---Altermagnets represent a recently identified class of magnetic materials characterized by collinear spin arrangements and unorthodox spin symmetry rules~\cite{PhysRevX.12.040501,PhysRevX.12.031042,PhysRevX.12.040002}. Unlike ferromagnets or antiferromagnets, they exhibit neither net magnetization nor conventional Néel order~\cite{PhysRevX.12.040501,PhysRevX.12.040002,PhysRevX.12.031042,doi:10.1126/sciadv.aaz8809,doi:10.1073/pnas.2108924118,Krempaský2024,PhysRevX.12.011028,PhysRevLett.130.216701,PhysRevLett.128.197202,PhysRevLett.134.196703,bzzy-ngcs}, but remain invariant under combined spin and specific crystal rotations~\cite{PhysRevX.12.040501}. This symmetry breaking yields nonrelativistic spin splitting, highly anisotropic spin transport, and distinctive magnetic responses despite vanishing total moment. These properties make altermagnets promising platforms for next-generation spintronic technologies~\cite{doi:10.1126/sciadv.aaz8809,doi:10.1073/pnas.2108924118,PhysRevX.12.011028,PhysRevLett.130.216701,PhysRevLett.128.197202,Feng2022,fu2025allelectricallycontrolledspintronicsaltermagnetic}.
		
		Their zero-moment ground state minimizes stray fields—desirable in low-noise and interference-sensitive applications—but hinders magnetic control through external fields~\cite{PhysRevX.12.040501,PhysRevX.12.040002,PhysRevX.12.031042,Smejkal2018,Krempaský2024}. Static methods to induce magnetization have been demonstrated~\cite{PhysRevLett.134.116701,PhysRevB.110.054446,br1r-bjzk,3r8w-76lf,golub2025spinorientationelectriccurrent}, yet dynamical generation remains open~\cite{PhysRevB.111.L140409,PhysRevB.111.134414,2rrr-glyn,fu2025floquetengineeringspintriplet}. Dynamical routes offer access to far-from-equilibrium regimes, enabling ultrafast and reversible spin control without altering intrinsic material properties.\begin{figure}[t]
			\centering
			\includegraphics[width=1\linewidth]{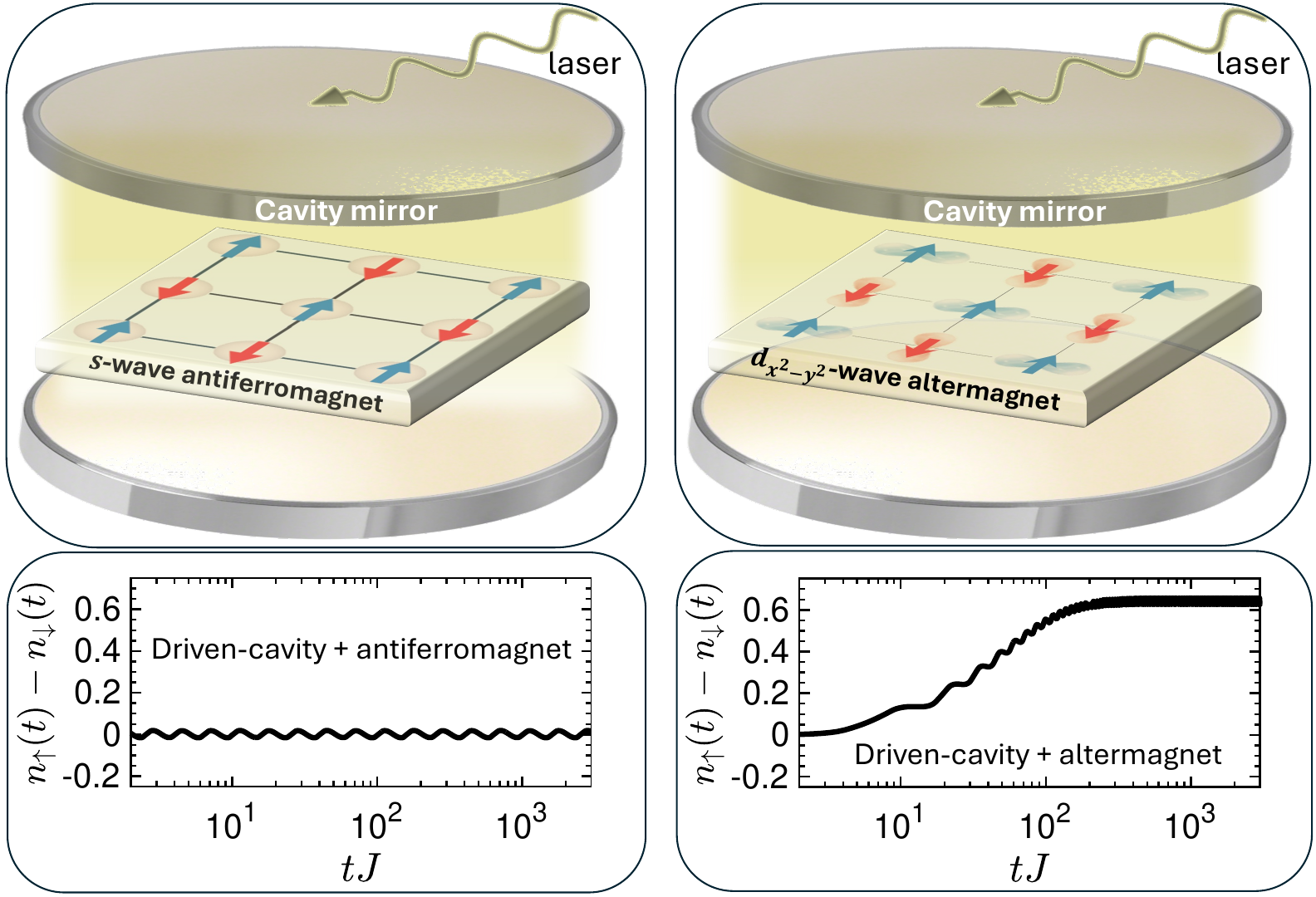}
			\caption{Schematic illustration of a laser-driven cavity setup hosting (left) a conventional $s$-wave antiferromagnet and (right) a $d_{x^2-y^2}$-wave altermagnet. The coherent laser drive excites cavity photons, which couple to spin textures via light-matter interaction. In the antiferromagnetic case, the spin populations remain nearly balanced under periodic driving, yielding minimal nonequilibrium polarization. In contrast, the anisotropic altermagnetic order due to staggered sublattices leads to a pronounced photoinduced imbalance between spin populations, as shown by the time evolution of the spin-resolved occupation difference $n_{\uparrow}(t)-n_{\downarrow}(t)$ (bottom panels).\iffalse (Left) Spin-resolved band structure of a $d_{x^2-y^2}$-wave altermagnet along a high-symmetry $\vec{k}$-path, showing momentum-dependent spin splitting between spin-up and spin-down bands. The inset depicts the collinear antiparallel spin arrangement characteristic of altermagnetic order. (Right) Schematic of the proposal: a $d_{x^2-y^2}$-wave altermagnet embedded in a coherently driven optical cavity. The coherent laser drive excites cavity photons, which couple to staggered sublattices and enable cavity-induced coherent magnetization.\fi}
			\label{f1}
		\end{figure}
		
		Ultrafast optical and cavity quantum electrodynamics techniques provide powerful tools for manipulating quantum materials~\cite{Lu25,Ebbesen2016Hybrid,Shalabney2015Coherent,Basov2025PolaritonicQuantumMatter,kass2024manybodyphotonblockadequantum,Yang2024ElectrochemicallySwitchableCavities,Bujalance2024,keren2025cavityalteredsuperconductivity,LeDe_2022,PhysRevB.111.155416,10.1063/5.0083825,doi:10.1126/sciadv.aau6969,Eckhardt2022,yarmohammadi2025cavityassistedmagnetizationswitchingquantum,ZARERAMESHTI20221,PhysRevA.94.033821,PhysRevX.11.031053,PhysRevApplied.19.064007}. In altermagnets, optical driving can reorient spins, tune orbitals, and reshape bands~\cite{Weber2024,Rajpurohit2024,Steward2023,PhysRevB.105.155138}. Strong light-matter coupling in optical cavities creates hybrid polaritonic states that reshape electronic, vibrational, or magnetic excitations~\cite{Ebbesen2016Hybrid,Ruggenthaler2018,FriskKockum2019,Hubener2021,Ruggenthaler2018,FriskKockum2019,Hubener2021,10.1063/PT.3.4749,10.1063/5.0083825,LeDe_2022,Li2018,PhysRevB.99.235156,kipp2024cavityelectrodynamicsvander,PhysRevLett.131.023601,PhysRevB.105.205424,QIN20241}. In magnetic systems, cavity photons can mediate long-range interactions and tune collective modes, as recently demonstrated for magnon-polaritons in altermagnets~\cite{jin2023cavityinducedstrongmagnonmagnoncoupling}. However, direct control of altermagnetic properties using driven cavities has not yet been addressed.
		
		In this Letter, we study a two-dimensional~(2D) $d$-wave altermagnet coupled to a coherently driven optical cavity. Unlike conventional antiferromagnets, where opposite sublattice responses cancel, sublattice-selective photon coupling in altermagnets breaks the compensation underlying the zero-moment state, producing a finite, tunable magnetization. Employing the mean-field Lindblad framework, we analyze both linear and quadratic electron-photon interactions and show that quadratic coupling is particularly effective, generating substantial magnetization at minimal coupling strengths. The resulting nonequilibrium steady state~(NESS) exhibits characteristic polariton signatures in the spin imbalance. Our findings establish a cavity-mediated route for dynamic magnetization control in altermagnets, opening pathways for ultrafast spintronic functionalities.
		
		\blue{\textit{Model.}}---We consider a 2D $d$-wave altermagnet described by a two-sublattice tight-binding model on a square lattice, where $\uparrow,\downarrow$ label spin degrees of freedom arising from sublattice distinctions. The model incorporates direction-dependent hopping (staggered fields), electron-photon interactions, and a driven cavity photon mode; see Fig.~\ref{f1}. In the absence of the cavity, the model captures the key features of $d$-wave altermagnetism. Candidate materials include KV$_2$Se$_2$O, RbV$_2$Te$_2$O, $\kappa$-Cl, RuO$_2$, and FeSe~\cite{Jiang2025,Zhang2025,Naka2019,Feng2022,doi:10.1126/sciadv.aaz8809,PhysRevLett.128.197202,weber2024opticalexcitationspinpolarization,PhysRevX.12.040501,C5CP07806G,Smejkal2022}, several of which have been experimentally confirmed to exhibit altermagnetic order and can be exfoliated to mimic the 2D geometry used here. The total Hamiltonian reads as $H = H_{\rm e}+H_{\rm ph}+H_{\rm e\text{-}ph}$, where $H_{\rm e}$ describes electrons, $H_{\rm ph}$ photons, and $H_{\rm e\text{-}ph}$ their coupling.
		
		The electronic part in real space is
		\begin{align}
			H_{\rm e} =& -J \sum_{\langle \ell,j\rangle,\sigma}\!\left(c_{\ell\sigma}^\dagger c_{j\sigma}+\text{h.c.}\right)\notag\\
			&+\sum_{\langle\!\langle \ell,j\rangle\!\rangle}\!\Delta_{\ell j}\left(c_{\ell\uparrow}^\dagger c_{j\uparrow}-c_{\ell\downarrow}^\dagger c_{j\downarrow}+\text{h.c.}\right),
		\end{align}
		where \( c_{\ell\sigma}^\dagger \) (\( c_{\ell\sigma} \)) creates (annihilates) an electron with spin \( \sigma = \uparrow, \downarrow \) at lattice site \(\ell\) on a square lattice with $N$ sites. The first term describes spin-independent nearest-neighbor hopping with amplitude \(J\), while $\Delta_{\ell j}=\pm\Delta_0$ along $x$/$y$ reflects the staggered altermagnetic order. The momentum-space form follows from a Fourier transform
		$c_{\ell\sigma} = \frac{1}{\sqrt{N}}\sum_{\vec{k}}e^{i\vec{k}\cdot\vec{r}_\ell}c_{\vec{k}\sigma}$: $H_{\text{e}} = \sum_{\vec{k}, \sigma} \epsilon_{\vec{k}} \, c_{\vec{k}\sigma}^\dagger c_{\vec{k}\sigma} + \sum_{\vec{k}} \Delta_{\vec{k}} \left( c_{\vec{k}\uparrow}^\dagger c_{\vec{k}\uparrow} - c_{\vec{k}\downarrow}^\dagger c_{\vec{k}\downarrow} \right)$ with
		\begin{subequations}
			\begin{align}
				\epsilon_{\vec{k}} &=-2J[\cos(k_x a)+\cos(k_y a)],\\
				\Delta_{\vec{k}} &=+2\Delta_0[\cos(k_x a)-\cos(k_y a)],
			\end{align}
		\end{subequations}
		and dispersion $\mathcal{E}_{\vec{k}}=\epsilon_{\vec{k}}\pm\Delta_{\vec{k}}$~\footnote{In our framework, we introduce the notion of spin magnetization using the Pauli matrix \(\sigma_z\) in orbital space, assigning \(\sigma_z = +1\) to the first orbital and \(\sigma_z = -1\) to the second. The total pseudospin magnetization per unit cell is defined as $n_\uparrow -n_\downarrow \propto  \int_{\mathrm{BZ}} \frac{d^2 k}{(2\pi)^2} \langle \sigma_z \rangle_{\vec{k}}$. Due to the antisymmetry of the spin-splitting term, \(\Delta_{(k_x,k_y)} = - \Delta_{(k_y,k_x)}\), under interchange of \(k_x\) and \(k_y\), and the corresponding symmetry of the Brillouin zone, the integral evaluates to zero. Consequently, in the absence of cavity photons, the momentum-dependent spin splitting does not produce a net spin magnetization, implying that spin densities remain perfectly balanced across the Brillouin zone, as shown in Fig.~\ref{f1} (left).}. The antisymmetry $\Delta_{(k_x,k_y)}=-\Delta_{(k_y,k_x)}$ enforces zero net magnetization in equilibrium~\cite{Feng2022,doi:10.1126/sciadv.aaz8809,PhysRevLett.128.197202,weber2024opticalexcitationspinpolarization,PhysRevX.12.040501,C5CP07806G,Smejkal2022,yarmohammadi2025anisotropiclighttailoredrkkyinteraction}. In the following, we demonstrate that this antisymmetry is disrupted once cavity photons are coherently driven and coupled to the spin-polarized electrons. The resulting photon-assisted processes generate a momentum-space imbalance in spin densities, thereby inducing a finite magnetization in the altermagnet.
		
		The cavity is modeled as a single driven mode in the dipole approximation~(long-wavelength optical cavity)~\cite{PhysRevB.111.085114}, given, in real space, by 
		\begin{equation}
			H_{\rm ph}=\omega_{\rm photon} a^\dagger a+\mathcal{E}_{\rm laser}\cos(\omega_{\rm laser} t)\sqrt{N}(a+a^\dagger),
		\end{equation}where \(a\) and \(a^\dagger\) denote the bosonic annihilation and creation operators for cavity photons, corresponding to a single mode with frequency \(\omega_{\rm photon}\) and coherent drive amplitude $\mathcal{E}_{\rm photon}$ with frequency \(\omega_{\rm laser}\)~\footnote{While laser pulses could also excite cavity photons and potentially induce magnetization over very short timescales, our focus here is on creating NESS, allowing the induced magnetization to be sustained and probed over extended timescales.}.
		
		Electron-photon coupling enters through the Peierls substitution~\cite{PhysRevLett.131.023601,kipp2024cavityelectrodynamicsvander,Eckhardt2022,Weber2023,Lu25}, $t_{ij}\to \tilde{t}_{ij} =  t_{ij}\exp\!\left[-\frac{ie}{\hbar}\int_{\vec{r}_j}^{\vec{r}_i}d\vec{r}\cdot\vec{A}(t)\right]$, where $\vec{A}(t)=\mathcal{A}_0(a+a^\dagger)\vec{\epsilon}$ is the cavity field with the polarization direction of $\vec{\epsilon}$  and the mode amplitude of $\mathcal{A}_0$. In altermagnets, both the nearest-neighbor hopping \(J\) and the spin-dependent next-nearest-neighbor term \(\Delta_0\) acquire the same Peierls phase. The resulting cavity-altermagnet Hamiltonian reads as $\widetilde{H}_{\rm e} = \sum_{\ell,j,\sigma} \, \tilde{t}_{ij}\, c_{\ell\sigma}^\dagger \left(- J \sigma_0 + \Delta_0 \sigma_z\right) c_{j\sigma}$, , where $\sigma_0~(\sigma_z)$ is the $2\times 2$ identity~(Pauli) matrix. The spin texture arises from the staggered potential, while photons couple orbitally via minimal coupling~\cite{PhysRevLett.132.236701}. This captures light-matter hybridization. Expanding to quadratic order, $e^{-\frac{ie}{\hbar}\vec{A}\cdot\vec{d}_{ij}}\!\approx 1-\frac{ie}{\hbar}\vec{A}\cdot\vec{d}_{ij}-\frac{e^2}{2\hbar^2}(\vec{A}\cdot\vec{d}_{ij})^2$ with $\vec{d}_{ij} = \vec{r}_\ell - \vec{r}_j$, reveals linear (paramagnetic) and quadratic (diamagnetic) terms. We retain the quadratic term, capturing orbital anisotropy that enables nonlinear electron-photon interactions. The factor of \(i\) arising in intermediate steps of the expansion is absorbed into the definition of the bond-current operator, so the interaction Hamiltonian remains manifestly Hermitian and free of spurious complex factors.
		
		Physically, the cavity couples more strongly to longer-range charge motion due to the greater phase accumulation. Because next-nearest-neighbor bonds ($\Delta_0$) are longer than nearest-neighbor ones ($J$), the light-matter coupling is enhanced for $\Delta_0$, which dominates long-wavelength cavity interactions. Projecting onto these bonds gives\begin{align}\label{eq_5}
			H_{\text{e-ph}} \approx {} &\Delta_0\sum_{\ell,j} \left[\lambda_{\rm linear}  \left(a + a^\dagger\right)+ \lambda_{\rm quadratic} \left(a + a^\dagger\right)^2\right] \notag \\ {} & \times \left(c_{\ell\uparrow}^\dagger c_{j\uparrow} - c_{\ell\downarrow}^\dagger c_{j\downarrow}\right),
		\end{align}with dimensionless couplings $\lambda_{\rm linear} = g_{\rm linear}/\Delta_0$ and $\lambda_{\rm quadratic} = g_{\rm quadratic}/\Delta_0$ set by bond geometry and polarization~\cite{PhysRevLett.131.023601,kipp2024cavityelectrodynamicsvander,Eckhardt2022,Weber2023,Lu25}. This local interaction term selectively couples photon displacement operators to the on-site electron density, while conserving spin. The staggered potential $\Delta_0$ breaks local inversion symmetry between sublattices, making the coupling spin selective. This photon-induced imbalance between spin sectors drives a finite pseudospin magnetization, as illustrated schematically in Fig.~\ref{f1}(right). In a conventional collinear antiferromagnet ($\Delta_0=0$) with the same lattice configuration, see Fig.~\ref{f1}(left), no such asymmetry exists, and cavity driving cannot break local inversion symmetry between sublattices on the same square lattice. In the thermodynamic limit, quantum fluctuations of the cavity field scale as $1/\sqrt{N}$ and can be neglected, enabling a mean-field treatment of $H_{\rm e\text{-}ph}$. 
		
		\blue{\textit{Lindbladian equations of motion}.}---Up to this point, the altermagnet-cavity system was treated as closed. Once the cavity is driven, energy is pumped into photons and electrons, necessitating dissipation to stabilize a NESS and prevent overheating. We model two baths: an internal bath, into which electrons and photons dissipate, and an external bath responsible for photon losses.
		
		Dissipation is described via the Markovian Lindblad master equation~\cite{Lindblad1976}:
		\begin{equation}
			\langle \dot{O} \rangle = i \langle [H,O]\rangle + \sum_i \gamma_i \Big(\langle L_i^\dagger O L_i \rangle - \tfrac{1}{2} \langle \{L_i^\dagger L_i,O\}\rangle\Big),
		\end{equation}
		with jump operators $L_i = \{c_{\vec{k}\sigma}, c_{\vec{k}\sigma}^\dagger, a, a^\dagger\}$ and damping rates $\gamma_i = \{\gamma_{\rm electron} (N_{\vec{k}}^\sigma-1), \gamma_{\rm electron} N_{\vec{k}}^\sigma, \gamma_{\rm ph} (N+1), \gamma_{\rm ph} N\}$. Here, $N_{\vec{k}}^\sigma$ and $N$ are fermionic and bosonic occupation numbers, and $\gamma_{\rm ph}  = \gamma_{\rm photon} $ or $\kappa$ accounts for both internal and external photon leakages. The total photon damping is then $\gamma_{\rm photon}+ \kappa$. 
		
		Within a mean-field treatment, the dynamics are captured by coupled equations for the cavity photon displacement $q = \langle a^\dagger + a \rangle/\sqrt{N}$, momentum $p = i \langle a^\dagger - a \rangle/\sqrt{N}$, and altermagnet electron densities $n_\sigma = \sum_{\vec k} \langle c^\dagger_{\vec{k}\sigma} c_{\vec{k}\sigma} \rangle/N$:
		\begin{subequations}
			\begin{align}
				&	\dot{q}(t) = \omega_{\rm photon} p(t) - \frac{\gamma_{\rm photon}+\kappa}{2} q(t)\,, \\
				&	\dot{p}(t) = - \omega_{\rm photon} q(t) - 2 \Big[\mathcal{E}(t) +\widetilde{g}(t) (n_\uparrow(t) - n_\downarrow(t) - 1) \Big] \notag \\ {} &\hspace*{0.95cm}- \frac{\gamma_{\rm photon}+\kappa}{2} p(t)\,, \\
				&	\dot{n}_\sigma(t) = - \gamma_{\rm electron} \Big(n_\sigma(t) - \frac{1}{N} \sum_{\vec{k}} N_{\vec{k}}^\sigma(t) \Big)\,,
			\end{align}
		\end{subequations}with $\widetilde{g}(t) = g_{\rm linear}+2 g_{\rm quadratic} q(t)$, $\beta = \frac{1}{k_{\mathrm{B}} T}$ being the inverse temperature, and instantaneous occupation numbers
		\begin{align}
			N_{\vec{k}}^\sigma(t) = \frac{1}{e^{\beta \omega_{\vec{k}}^\sigma(t)}+1},\quad 
			\omega_{\vec{k}}^\sigma(t) = \epsilon_{\vec{k}} + \sigma \big(\Delta_{\vec{k}} + \widetilde{g}(t) q(t)\big),
		\end{align}where the interplay between electrons and photons manifests in the distribution function instead of in the bare electronic energy spectrum. 
		
		We focus on the strong altermagnetic limit $\Delta_0 \approx J$, where the spin splitting is comparable to the bandwidth, allowing analytical insight. For weaker $\Delta_0$, qualitative behavior remains similar but requires numerical evaluation. Independent of the $\Delta_0 > 0$ value, the net magnetization remains zero in the absence of photons. In the strong altermagnetic regime, where $\Delta_0 \sim J$, an analytical expression for the NESS spin densities ($\dot{n}_{\sigma}(t)=0$) can be derived in the zero-temperature limit ($T\!\to\!0$):  
		\begin{equation}
			\frac{1}{N} \sum_{\vec{k}} N_{\vec{k}}^{\sigma}(t)
			\simeq
			\frac{1}{\pi} \arccos\!\left(\frac{\sigma\, \widetilde{g}(t)\, q(t)}{4J}\right).
			\label{eq:NESS_spin_density}
		\end{equation}
		This relation explicitly shows that the cavity-induced photon displacement $q(t)$ directly controls the fraction of occupied electronic states for each spin species, thereby generating a spin imbalance and, consequently, a cavity-induced magnetization.
		\begin{figure}[t]
			\centering
			\includegraphics[width=1\linewidth]{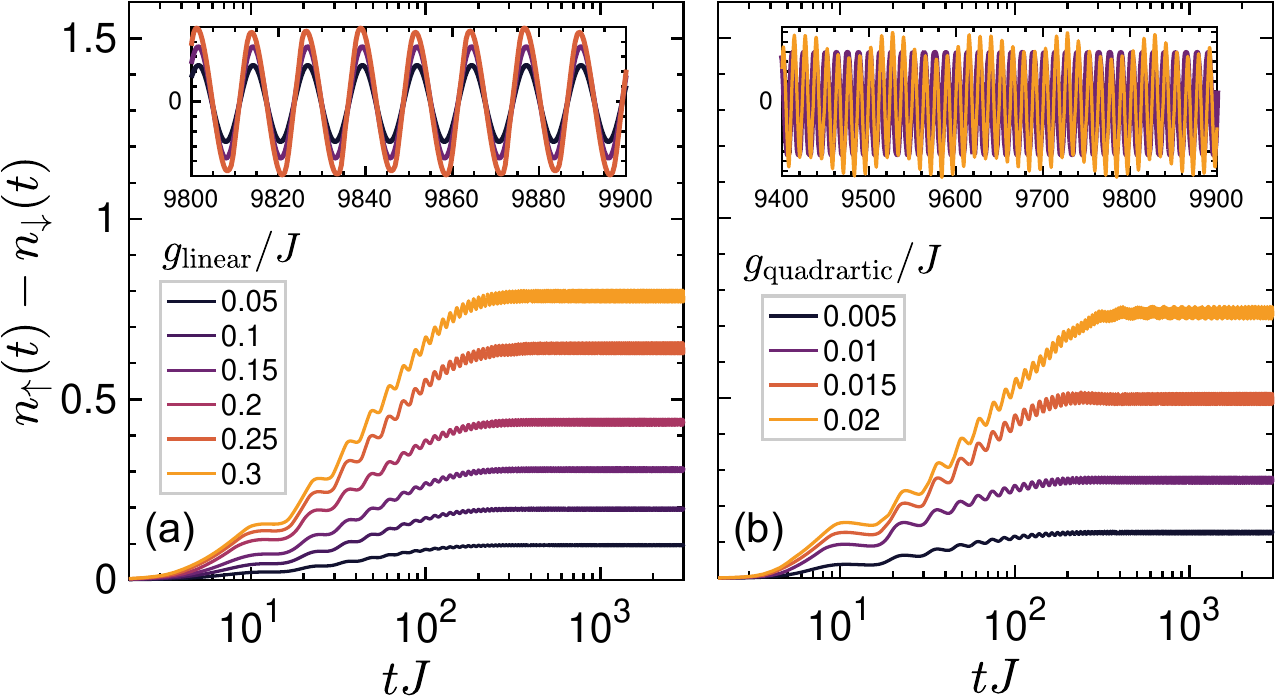}
			\caption{Time evolution of the spin imbalance $n_{\uparrow}(t)-n_{\downarrow}(t)$ for (a) varying $g_{\rm linear}$ with $g_{\rm quadratic}=0$ and (b) varying $g_{\rm quadratic}$ with $g_{\rm linear}=0$. All couplings reach a NESS, with stronger couplings accelerating the onset and enhancing the steady-state magnetization, reflecting cavity-induced sublattice symmetry breaking. Insets display coherent long-time oscillations, characteristic of the NESS, for several coupling strengths, obtained after subtracting their respective NESS values to align all traces around zero for comparison. Parameters: $\omega_{\rm photon}/J=0.5$, $\omega_{\rm laser}=\omega_{\rm photon}$, $\mathcal{E}_{\rm laser}/J=0.1$, $\gamma_{\rm electron}=\gamma_{\rm photon}=\kappa=0.02J$.}
			\label{f2}
		\end{figure}
		
		\blue{\textit{Results and discussion.}}---In our simulations, parameters are chosen such that $\left| \widetilde{g}(t) q(t)\right| < 4J$, ensuring sublattice symmetry breaking and a finite spin-density imbalance (magnetization). All energies are expressed in units of $J = 1$, with $\max(g_{\rm linear}) = 0.6~\mathrm{eV}$, $\max(g_{\rm quadratic}) = 30~\mathrm{meV}$, $\omega_{\rm laser} \le 1~\mathrm{eV}$, $\omega_{\rm photon} \le 1~\mathrm{eV}$, and damping rates $\le 60~\mathrm{meV}$. These ensure that the spin population imbalance does not exceed unity while establishing the steady-state, thereby avoiding overheating or unphysical regimes.
		
		Figure~\ref{f2} shows the time evolution of the spin imbalance $n_{\uparrow}(t) - n_{\downarrow}(t)$ for various (a) $g_{\rm linear}/J$ and (b) $g_{\rm quadratic}/J$, at $\omega_{\rm laser} = \omega_{\rm photon} = 0.5\,J$. At early times ($tJ \lesssim 5$), all curves remain close to zero, indicating that sublattice symmetry is largely preserved. At intermediate times ($5 \lesssim tJ \lesssim 300$), the imbalance grows monotonically with $g$, demonstrating efficient cavity-induced sublattice symmetry breaking. At late times ($tJ \gtrsim 300$), the system reaches a long-lived steady plateau, indicating the formation of a coherent magnetization; for $J = 1~\mathrm{eV}$, this corresponds to $\sim 200$~fs. The saturation value increases with coupling strength, with $g_{\rm linear}/J = 0.3$ and $g_{\rm quadratic}/J = 0.02$ producing $\sim 80\%$ imbalance. The inset oscillations on the long-time plateau reflect residual coherent dynamics in the NESS. Thus, coupling the 2D \(d_{x^2-y^2}\)-wave altermagnet to the cavity field induces a finite coherent magnetization by breaking sublattice symmetry and creating a spin population imbalance. \begin{figure}[t]
			\centering
			\includegraphics[width=0.9\linewidth]{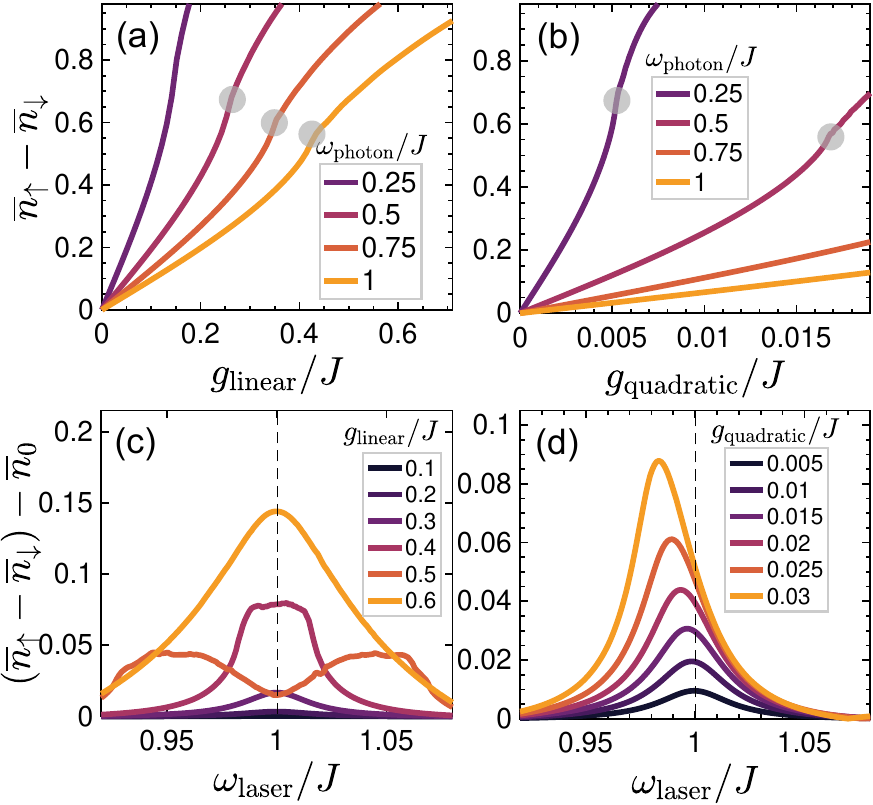}
			\caption{Time-averaged cavity-induced magnetization in the NESS, $\overline{n}_{\uparrow}-\overline{n}_{\downarrow}$, as a function of (a) $g_{\rm linear}/J$ at $g_{\rm quadratic}=0$ and (b) $g_{\rm quadratic}/J$ at $g_{\rm linear}=0$, for various $\omega_{\rm photon}/J$ at $\omega_{\rm laser}=\omega_{\rm photon}$. At weak coupling, the response is perturbative, while above a critical coupling, it becomes strongly nonlinear due to polariton formation; see small gray circles and (c,d). Nonlinear enhancement is more pronounced for linear coupling at lower photon frequencies. (c,d) Magnetization versus $\omega_{\rm laser}/J$ at $\omega_{\rm photon}/J=1$, for different (c) linear and (d) quadratic couplings. Baseline magnetization $\overline{n}_0$ at $\omega_{\rm laser}/J=0.9$ is subtracted. A resonant peak near $\omega_{\rm laser}\simeq\omega_{\rm photon}$ appears for all $g_{\rm linear}$, except for $0.4<g_{\rm linear}/J<0.55$, where symmetric double peaks appear, signaling polariton formation, consistent with the position of small gray dot in (a). Parameters: $\mathcal{E}_{\rm laser}/J=0.1$ and $\gamma_{\rm electron}=\gamma_{\rm photon}=\kappa=0.02J$. 
			}
			\label{f3}
		\end{figure}
		
		Figure~\ref{f3} shows the NESS spin imbalance $\overline{n}_{\uparrow}-\overline{n}_{\downarrow}$ as a function of (a) linear coupling $g_{\rm linear}/J$ and (b) quadratic coupling $g_{\rm quadratic}/J$, for photon frequencies $\omega_{\rm photon}/J = \{0.25,\,0.5,\,0.75,\,1\}$. We consider the resonant regime, where the laser frequency matches the cavity photon frequency, $\omega_{\rm laser} = \omega_{\rm photon}$. The NESS magnetization is obtained by time-averaging over one oscillation period after evolving to $tJ \approx 10^5$, i.e., to $\sim 66$ ps. In both coupling schemes, the spin imbalance grows monotonically with coupling, with stronger couplings yielding larger cavity-induced magnetization. Quadratic coupling is particularly effective: even at $g_{\rm quadratic}/J \sim 0.02$, it generates substantial magnetization, while linear coupling requires $g_{\rm linear}/J \gtrsim 0.5$ to achieve comparable effects. The response is enhanced at lower photon frequencies due to the larger steady-state photon displacement, $\overline{q}_{\rm NESS} \propto 2 \mathcal{E}_{\rm laser}/(\gamma_{\rm photon}\,\omega_{\rm photon})$.
		
		For couplings below a critical value $g^c$, the magnetization scales perturbatively as $g^2$, a consequence of the symmetry $g \mapsto -g$ (or equivalently $q \mapsto -q$). Increasing $g$ drives a crossover to a nonlinear regime, where the magnetization grows faster than quadratic and the system exhibits novel features. In particular, for linear coupling at $\omega_{\rm photon}/J = 1$, a symmetric double-peak structure emerges in the laser-frequency response in Fig.~\ref{f3}(c) for $0.4 \lesssim g_{\rm linear}/J \lesssim 0.55$, signaling polariton formation from coherent hybridization of cavity photons with electronic excitations~\cite{Xiang2024_MolecularPolaritons,Schwennicke2024_DisorderedPolaritons,GarciaVidal2021_ManipulatingMatter,Luo2024,Tay2025,SM}. The peak separation approximates the Rabi frequency, reflecting vacuum Rabi splitting. Note that from the peak positions, one can identify the dispersions of the upper and lower hybridized modes. Quadratic coupling shows no clear polariton signature at this photon frequency, though the effective photon frequency is renormalized by the induced spin imbalance, $\widetilde{\omega}_{\rm photon} = \omega_{\rm photon} \sqrt{1 + 4 \frac{g_{\rm quadratic}}{\omega_{\rm photon}} (\overline{n}_{\uparrow} - \overline{n}_{\downarrow})}$, shifting the resonance as $g_{\rm quadratic}$ increases.\begin{figure}[t]
			\centering
			\includegraphics[width=0.9\linewidth]{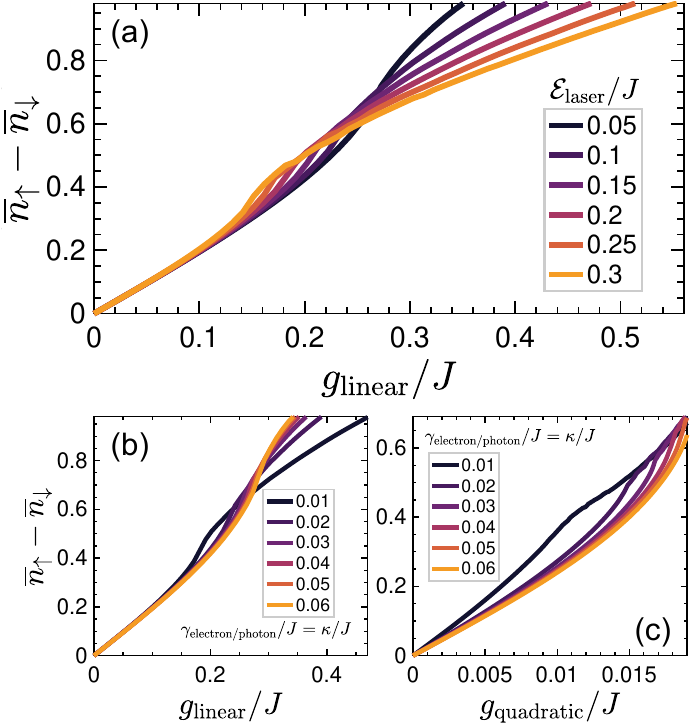}
			\caption{(a) Cavity-induced coherent magnetization versus linear electron-photon coupling $g_{\rm linear}/J$ for several laser amplitudes $\mathcal{E}_{\rm laser}/J$. Larger drive amplitudes extend the nonlinear regime, shifting the onset of polariton formation to lower $g_{\rm linear}/J$. 
				Steady-state spin imbalance $\overline{n}_{\uparrow} - \overline{n}_{\downarrow}$ as a function of coupling strength for (b) purely linear ($g_{\rm quadratic}/J = 0$) and (c) purely quadratic ($g_{\rm linear}/J = 0$) couplings, shown for different equal dissipation rates $\gamma_{\rm electron} = \gamma_{\rm photon} = \kappa$. In the linear case (b), the imbalance transitions from quadratic scaling at weak coupling to a steeper, nonlinear increase beyond a damping-dependent threshold. In the quadratic case (c), the response remains nearly quadratic but exhibits curvature changes at higher coupling. Parameters in (a): $\gamma_{\rm electron} = \gamma_{\rm photon} = \kappa = 0.02 J$ and $g_{\rm quadratic}/J = 0$; in (b): $\mathcal{E}_{\rm laser}/J = 0.1$. Fixed parameters: $\omega_{\rm photon}/J = 0.5$, $\omega_{\rm laser} = \omega_{\rm photon}$.
			}
			\label{f5}
		\end{figure} 
		
		The video provided in the SM~\cite{SM} confirms the formation of polaritons induced by linear coupling for various laser amplitudes at \(\omega_{\rm photon}/J = 1\). The blue curves in the video display normal resonant single peaks for \(\mathcal{E}_{\rm laser}/J = 0.05\) and 0.1, whereas the red double peaks signify polaritons for \(\mathcal{E}_{\rm laser}/J > 0.1\). 
		
		Thus, these results demonstrate that cavity driving not only induces a finite NESS magnetization in altermagnets through sublattice-symmetry breaking, but also allows the formation of polaritonic excitations through strong linear coupling, highlighting the interplay of cavity photons and spin-charge degrees of freedom in the NESS. As discussed before, we focus on the strong altermagnetic limit, $\Delta_0 \sim J$, where we find analytical expressions for the steady-state spin densities, capturing the essential physics of cavity-induced magnetization. We stress that detailed polariton behavior for other $\Delta_0$ values, however, depends sensitively on cavity frequency, light-matter coupling, and band structure.
		
		Figure~\ref{f5}(a) shows the effect of the laser amplitude on cavity-induced magnetization versus the linear electron-photon coupling \(g_{\rm linear}/J\) for $g_{\rm quadratic}/J = 0$, $\omega_{\rm photon}/J = 0.5$, $\omega_{\rm laser} = \omega_{\rm photon}$, and $\gamma_{\rm electron} = \gamma_{\rm photon} = \kappa = 0.02 J$. The magnetization initially grows quadratically with $g_{\rm linear}$, up to a laser-dependent threshold marking polariton formation. Beyond this threshold, the response enters a nonlinear regime, with stronger drives shifting the onset to lower $g_{\rm linear}$, reflecting a cooperative enhancement of light-matter hybridization and spin imbalance. The laser field thus enhances the spin-symmetry-breaking channels, which indicates that experimentally accessible laser amplitudes can relax material constraints on intrinsic coupling strengths. Similar trends are expected for quadratic couplings, with appropriately shifted thresholds.
		
		Finally, Figs.~\ref{f5}(b,c) display the NESS spin imbalance versus coupling strength for equal damping rates $\gamma_{\rm electron} = \gamma_{\rm photon} = \kappa$ varied from $0.01J$ to $0.06J$. For linear coupling, small $g_{\rm linear}$ shows quadratic growth, with the nonlinear crossover occurring earlier at lower damping. Higher dissipation smooths this transition and suppresses the peak imbalance. For quadratic coupling, the imbalance increases smoothly over the full range, with damping modulating the curvature and growth rate, indicating weaker sensitivity to cooperative nonlinear effects but persistent damping dependence.
		
		\blue{\textit{Experimental perspective.}}---The proposed cavity-induced magnetization in altermagnets can be probed using current platforms combining high-quality optical cavities with thin-film or exfoliated materials, such as KV$_2$Se$_2$O, RbV$_2$Te$_2$O, RuO$_2$~(it remains under debate), and $\kappa$-Cl~\cite{Jiang2025,Zhang2025,Feng2022,doi:10.1126/sciadv.aaz8809,PhysRevLett.128.197202,weber2024opticalexcitationspinpolarization,PhysRevX.12.040501,C5CP07806G,Smejkal2022,Naka2019}, which exhibit $d$-wave spin textures and momentum-dependent spin splitting. Embedding these samples in Fabry-Pérot or photonic-crystal cavities~\cite{Ebbesen2016, Sentef2022} and driving with a tunable continuous-wave laser~\footnote{Since our proposal relies on a weak continuous-wave laser and weak cavity-altermagnet coupling, the energy input remains well below the lattice-melting threshold and is balanced by dissipation, ensuring a stable NESS.} enables control of $g_{\rm linear}$, $g_{\rm quadratic}$, and $\mathcal{E}_{\rm laser}$. 
		
		The resulting spin-density imbalance corresponds to a magnetic moment per site $(n_\uparrow - n_\downarrow)\mu_{\rm B}$, where $\mu_{\rm B}$ is the Bohr magneton. In spintronics, charge-spin conversion efficiency scales with the induced magnetic moment. Although the cavity-induced moments in our setup remain below $1\,\mu_{\rm B}$ per site, they are well within detection limits using spin-resolved ARPES or MOKE~\cite{10.1063/5.0151859,RevModPhys.82.2731}, both of which can probe steady-state magnetization under continuous illumination. Although electrons striking the cavity walls could pose a challenge for spin-ARPES, experimental setups can accommodate electron escape using partially reflective mirrors or small apertures.  
		
		Polaritons could also be detected via splittings in optical or electronic spectra as functions of coupling and detuning. By tuning damping rates through mirror coatings or sample-environment coupling, the predicted crossover from perturbative to nonlinear regimes and the cooperative enhancement between drive and intrinsic electron-photon coupling can be directly tested.
		
		\blue{\textit{Summary and outlook.}}---We show that embedding a two-dimensional $d$-wave altermagnet in a laser-driven optical cavity enables finite magnetization via light-matter interactions. Unlike conventional antiferromagnets, where sublattice cancellation suppresses net magnetization for the same lattice configuration, altermagnets’ symmetry-broken spin textures allow spin-resolved electron-photon coupling to generate a controllable spin imbalance even without spin-orbit coupling. Using tailored driving and dissipation within a mean-field Lindblad framework, we establish a non-equilibrium steady state where quadratic electron-photon coupling dominates, producing pronounced magnetization at minimal coupling. Strong coupling induces polaritons, leaving a distinctive signature in the steady-state spin imbalance, directly linking the optical cavity to spintronic functionality. 
		
		Our results highlight cavity engineering as a route to tunable, light-induced magnetic states in altermagnets. Further control could be achieved by coupling photons to lattice vibrations~\cite{88tw-h78r} or using frequency-chirped drives~\cite{PhysRevB.109.224417}, enabling magnetization switching and finer manipulation of spin currents. These strategies offer a promising platform for optically controlled spintronics and novel device concepts. We leave these topics to be explored in future work.
		
		\blue{\textit{Acknowledgments.}}---M.~Y.\ gratefully acknowledges the hospitality of the Max Planck Institute for the Physics of Complex Systems during his visit, where parts of this work were carried out. M. Y. and J. K. F. were supported by the Department of Energy, Office of Basic Energy Sciences, Division of Materials Sciences and Engineering under Contract No. DE-FG02-08ER46542 for the formal developments, the numerical work, and the writing of the manuscript. J. K. F. was also supported by the McDevitt bequest at Georgetown University. L. Š.\ acknowledges funding from the ERC Starting Grant No. 101165122.
		
		\blue{\textit{Data availability.}}--- The data that support the findings of this article are openly available at~\cite{Zenodo}.
	}
	\bibliography{Refs_Cavity_altermagnet_2025.bib}
	
\end{document}